\documentclass[referee]{raa}            

\usepackage{graphicx,times}             
\usepackage{natbib}
\usepackage{amssymb,amsmath}
\bibpunct{(}{)}{;}{a}{}{,}

\usepackage[a4paper=true,dvipdfm=true,pagebackref=true]{hyperref}
\hypersetup{colorlinks = true, linkcolor = green, anchorcolor = red, citecolor = blue, filecolor = red, pagecolor = red, urlcolor = red}

\usepackage{lscape,longtable}

\begin{document}

   \title{Extreme space weather events caused by super active regions during solar cycles 21-24}

 \volnopage{ {\bf 20XX} Vol.\ {\bf X} No. {\bf XX}, 000--000}
   \setcounter{page}{1}

   \author{Gui-Ming Le\inst{1,2}, Gui-Ang Liu\inst{1}, Ming-Xian Zhao\inst{2},
     Tian Mao\inst{2}, Ping-Guo Xu\inst{3}
   }

   \institute{School of Physics Science and Technology,
              Lingnan Normal University,
              Zhanjiang 524048, P.R. China; ({\it Legm@cma.gov.cn, zhaomx@cma.gov.cn, jdtpingguo@buu.edu.cn})\\
         \and
              Key Laboratory of Space Weather,
              National Center for Space Weather,
              China Meteorological Administration,
              Beijing, 100081, P.R. China\\
         \and
              Beijing Engineering Research Center of Smart Mechanical Innovation Design Service, Beijing, 100020, P.R. China\\
\vs \no
   {\small Received 20** June **; accepted 20** July **}
}

\abstract {Extreme space weather events including $\ge$X5.0 flares, ground level enhancement (GLE) events and super geomagnetic storms (Dst $\le$ -250 nT) caused by super active regions (SARs) during solar cycles 21-24 were studied. The total number of $\ge$X5.0 solar flares was 62, 41 of them were X5.0-X9.9 flares and 21 of them were $\ge$X10.0 flares. We found that 83.9\% of the $\ge$X5.0 flares were produced by SARs. 78.05\% of the X5.0-X9.9 and 95.24\% of the $\ge$X10.0 solar flares were produced by SARs. 46 GLEs registered during solar cycles 21-24, and 25 GLEs were caused by SARs, indicating that 54.3\% of the GLEs were caused by SARs. 24 super geomagnetic storms were recorded during solar cycles 21-24, and 12 of them were caused by SARs, namely 50\% of the super geomagnetic storms are caused by SARs. It is found that only 29 SARs can produce $\ge$X5.0 flares, 15 SARs can produce GLEs and 10 SARs can produce super geomagnetic storms. Of the 51 SARs, only 33 SARs can produce at least one extreme space weather event, while none of the rest 18 SARs can produce an extreme space weather event. There were only 4 SARs, each of them can produce not only  a $\ge$X5.0 flare, but also a GLE event and a super geomagnetic storm. Most of the extreme space weather events caused by the SARs appeared during solar cycles 22 and 23, especially for GLE events and super geomagnetic storms. The longitudinal distributions of source locations for the extreme space weather events caused by SARs were also studied.
}

   \authorrunning{G.-M. Le et al.}            
   \titlerunning{Extreme space weather and super geomagnetic storms}  
   \maketitle

%
\section{Introduction}
\label{sect:intro}

A major solar flare may lead to a sudden ionospheric disturbance, which may lead to sudden cosmic noise absorption induced by sudden electron density enhancement in the D region, short-wave fadeouts, sudden phase anomalies, sudden frequency disturbances, and a sudden increase in TEC \citep{Mendillo+etal+1974}. The duration of the effect of a solar flare on ionosphere ranges from several minutes to tens of minutes. The duration of a geomagnetic storm is much longer than a solar flare. The article by \cite{Richardson+etal+2006} found that largest geomagnetic storm caused by corotating interaction regions is weaker than a great geomagnetic storm (Dst $\le$ -200 nT) based on Burton equation \citep{Burton+etal+1975}, implying that a great geomagnetic storm can only be caused by associated coronal mass ejection (CME). Each solar cycle usually has about 3000 ARs. However, only small part of the ARs can produce very strong eruptions. These ARs are defined as super active regions (SARs). Many articles have been devoted to the study of the concept of SAR \citep[e.g.][and reference therein]{Bai+1987, Chen+etal+2011}. The flare that was accompanied by hard X-ray with peak flux $\ge$1000 counts s$^{-1}$ was defined as a major flare. If an active region (AR) can produce five or more major flares, then the AR was a SAR \citep{Bai+1987}. It is evident that \cite{Bai+1987} only linked SARs with solar flares. The definition of SAR proposed by \cite{Wu+Zhang+1995} is decided by five parameters: the largest area of the AR, the flare index of the X class X-ray flares, the peak flux of 10.7 cm radio flux, the short-term total solar irradiance decrease, and the peak flux of E$>$10 MeV protons. The concept of SAR proposed by \cite{Wu+Zhang+1995} linked SARs with both flare and solar proton events. However, the limited conditions for the five parameters has not been mentioned. Five parameters that were used to determine whether a AR is a SAR proposed by \cite{Tian+etal+2002} were: the largest area of the AR $\ge$1000 $\mu$h (millionths of solar hemisphere ), the flare index $\ge$5.0, the peak flux of 10.7cm radio flux $\ge$1000 s.f.u., and the Ap index $\ge$50, respectively. If a AR can satisfy three of the five parameters, then the AR is a SAR. The concept of SAR proposed by \cite{Tian+etal+2002} linked SARs with SPEs and geomagnetic storms. \cite{Romano+Zuccarello+2007} defined flare index as:
\begin{equation}\label{eq-01}
  I(t)=0.1 \times\sum{B(t)}+\sum{C(t)}+10 \times \sum{M(t)}+100 \times \sum{X(t)}
\end{equation}
where B(t), C(t), M(t) and X(t) are the coefficients of the flare that occurred at the time t and belonging to the class B, C, M and X, respectively.

If I(t) produced by a AR is greater than 500, then the AR is called as SAR by \cite{Romano+Zuccarello+2007}. It is evident that SARs defined by \cite{Romano+Zuccarello+2007} only linked with solar flares. Different researchers have different criteria for SARs, leading to different lists of SARs for same period. \cite{Chen+etal+2011} select an adequate set of criterion parameters and reparameterize the SARs during solar cycles 21-23. The parameters used to define SARs in the article by \cite{Chen+etal+2011} are: (1)the largest area of the AR is greater than 1000 $\mu$h, (2) flare index is larger then 10. Note that 0.1 for an M1 class flare and 1.0 for an X1 class flare in the calculation of the flare index proposed by \cite{Chen+etal+2011}, (3)The peak value of 10.7 cm radio flux$>$1000 s.f.u., (4) the short term total solar irradiance decrease ($\Delta$TSI) lower than 0.1\%. If a AR satisfy three of the four criterion conditions, then the AR is a SAR. If the flare index of a AR is larger than 15, and any one of the rest three other criterion conditions is met, then the AR is also a SAR. The criteria used by \cite{Chen+etal+2011} to select SARs have three properties. First, the parameters selected to determine SAR are independent, each providing a complementary insight into SAR physics. Secondly, the parameters can be easy to access. Thirdly, the number of parameters used to select SARs is both simple and unique.

It has been found that 44\% of the all X class X-ray flares during solar cycles 21–23 were produced by 45 SARs \citep{Chen+etal+2011}. However, little attention has been paid to the relationship between SARs and super geomagnetic storms (Dst $\le$ -250 nT) and GLE events. When a SAR erupts, it may only produce a flare, or it may produce both a flare and a CME, which may lead to a relativistic solar proton event and then causes a GLE event. If the CME and CME-driven shock finally reaches the Earth, it may trigger a super geomagnetic storm. In this context, SARs not only may produce very strong flares, but also may produce GLE events and super magnetic storms. There were 45 SARs during solar cycles 21-23, 5 SARs that appeared in solar cycle 24 were identified by \cite{Chen+Wang+2015} and AR 12673 is also a SAR according to the criteria proposed by \cite{Chen+etal+2011}. Hence, there were totally 51 SARs during solar cycles 21-24. The extreme space weather events are defined as solar flares with intensities $\ge$X5.0, super geomagnetic storms (Dst $\le$ -250 nT) and GLE events in this study. Because only one complete list of SARs during solar cycles 21-24 according to the criteria \citep{Chen+etal+2011} can be available at present. Now the question is how many extreme space weather events were caused by the SARs during solar cycles 21-24? To answer these questions, the extreme space weather events caused by the SARs during solar cycles 21-24 will be investigated based on the 51 SARs according to criteria proposed by \cite{Chen+etal+2011}. This is the motivation of the present study. The data analysis is presented in Section \ref{sect:data}, Discussion in Section \ref{sect:discussion} and the summary in Section \ref{sect:summary}.
\section{Data analysis}
\label{sect:data}
\subsection{Data Source}
\label{subsect:datasource}

The flares with intensities $\ge$X5.0 during solar cycles 21-24 were obtained from the website at
\url{ftp://ftp.ngdc.noaa.gov/STP/space-weather/solar-data/solar-features/solar-flares/x-rays/goes/xrs/}. The fluxes of E$>$10, 30, 50 and 100 MeV protons observed by GOES can be available from the website at \url{https://satdat.ngdc.noaa.gov/sem/goes/data/avg/}. The super geomagnetic storms were obtained from the website at \url{http://wdc.kugi.kyoto-u.ac.jp/dstdir/}. GLEs can be directly obtained from the appendix in the article by \cite{Le+Liu+2020}.

\subsection{$\ge$ X5.0 solar flares caused by SARs}
\label{subsect:dataX5}

According to the source locations of the ARs that produced $\ge$X5.0 flares and the list of 51 SARs, $\ge$X5.0 flares and the corresponding ARs are listed in Table \ref{tab:01}. In the Table, the number of the $\ge$X5.0 flares in column 1, the date of the flare in column 2, the start, peak and the end time of the flare in column 3, 4 and 5, respectively, the flare intensity in column 6, the source location of the flare in column 7, the NOAA number of AR in column 8 and whether the AR is a SAR in column 9. There were 62 flares with intensities $\ge$X5.0 during solar cycles 21-24 shown in Table \ref{tab:01}. Of the 62 $\ge$X5.0 flares, 9 of them were not caused by SARs, the source location for one flare that occurred on 4 March 1991 is unknown. 51 flares with intensities $\ge$X5.0 were caused by SARs, indicating that 83.9\% of the flares with intensities $\ge$X5.0 were caused by SARs. If we divide $\ge$X5.0 flares into two subgroups, the numbers of X5.0-X9.9 and $\ge$X10.0 flares are 41 and 21, respectively. It is derived that 78.05\% of the X5.0-X9.9 and 95.24\% of the $\ge$X10.0 flares were produced by SARs.

\begin{center}
\begin{longtable}{ccccccccc}
\caption[]{The flares with intensities $\ge$X5.0 caused by SARs during solar cycles 21-24.}\label{tab:01} \\
  \hline
  No. & Date & Start & Peak & End & Flare & Location & AR & SAR? \\
      &~yyyy-mm-dd~ & ~hh:mm~ & ~hh:mm~ & ~hh:mm~ & ~intensity~ & & & \\
  \hline
\endfirsthead
\multicolumn{9}{l}%
{{continued from previous page}} \\
  \hline
  No. & Date & Start & Peak & End & Flare & Location & AR & SAR? \\
      &~yyyy-mm-dd~ & ~hh:mm~ & ~hh:mm~ & ~hh:mm~ & ~intensity~ & & & \\
  \hline
\endhead

\hline \multicolumn{9}{r}{{continued on next page}} \\
\endfoot
\hline
\endlastfoot

1 & 1978-04-28 & 13:08 & 13:29 & 19:13 & X5.0 & N22E41 & 1092 & Yes \\
2 & 1979-08-18 & 14:03 & 14:16 & 14:45 & X6.0 & N10E90 & 1943 & No \\
3 & 1979-08-20 & 09:06 & 09:23 & 10:05 & X5.0 & N05E76 & 1943 & No \\
4 & 1979-09-19 & 22:56 & 23:03 & 23:45 & X5.0 & N06E33 & 1994 & No \\
5 & 1980-04-04 & 14:57 & 15:09 & 17:29 & X5.0 & N24W34 & 2363 & No \\
6 & 1980-11-06 & 03:40 & 03:48 & 04:48 & X9.0 & S12E74 & 2779 & Yes \\
7 & 1981-04-24 & 13:46 & 14:11 & 16:15 & X5.9 & N18W50 & 3049 & Yes \\
8 & 1981-04-27 & 07:20 & 08:20 & 09:45 & X5.5 & N17W90 & 3049 & Yes \\
9 & 1982-06-03 & 11:41 & 11:48 & 12:46 & X8.0 & S09E72 & 3763 & Yes \\
10 & 1982-06-04 & 13:13 & 13:30 & 13:58 & X5.9 & S10E55 & 3763 & Yes \\
11 & 1982-06-06 & 16:30 & 16:54 & 18:32 & X12.0 & S09E25 & 3763 & Yes \\
12 & 1982-07-09 & 07:31 & 07:38 & 08:24 & X9.8 & N17E73 & 3804 & Yes \\
13 & 1982-07-12 & 09:16 & 09:18 & 12:00 & X7.1 & N11E37 & 3804 & Yes \\
14 & 1982-12-15 & 01:50 & 01:59 & 02:46 & X12.9 & S10E24 & 4026 & Yes \\
15 & 1982-12-15 & 16:20 & 16:37 & 17:09 & X5.0 & S10E15 & 4026 & Yes \\
16 & 1982-12-17 & 18:19 & 18:58 & 20:23 & X10.1 & S08W21 & 4025 & No \\
17 & 1984-04-24 & 23:56 & 24:01 & 24:60 & X13.0 & S12E43 & 4474 & Yes \\
18 & 1984-05-20 & 22:18 & 22:41 & 23:57 & X10.1 & S09E52 & 4492 & Yes \\
19 & 1988-06-24 & 16:44 & 16:48 & 17:38 & X5.6 & S17W56 & 5047 & No \\
20 & 1989-03-06 & 13:54 & 14:10 & 15:04 & X15.0 & N35E69 & 5395 & Yes \\
21 & 1989-03-17 & 17:29 & 17:37 & 18:52 & X6.5 & N33W60 & 5395 & Yes \\
22 & 1989-08-16 & 01:08 & 01:17 & 02:28 & X20.0 & S18W84 & 5629 & Yes \\
23 & 1989-09-29 & 10:47 & 10:93 & 13:15 & X9.8 & S20W90 & 5698 & Yes \\
24 & 1989-10-19 & 12:29 & 12:55 & 17:33 & X13.0 & S27E10 & 5747 & Yes \\
25 & 1989-10-24 & 17:36 & 18:31 & 23:04 & X5.7 & S30W57 & 5747 & Yes \\
26 & 1990-05-21 & 22:12 & 22:17 & 23:39 & X5.5 & N35W36 & 6063 & Yes \\
27 & 1990-05-24 & 20:46 & 20:49 & 21:05 & X9.3 & N33W78 & 6063 & Yes \\
28 & 1991-01-25 & 06:30 & 06:30 & 06:38 & X10.0 & S16E78 & 6471 & Yes \\
29 & 1991-03-04 & 13:56 & 14:03 & 15:08 & X7.1 & unknown & unknown & unknown \\
30 & 1991-03-07 & 06:11 & 07:08 & 08:17 & X5.5 & S20E66 & 6538 & Yes \\
31 & 1991-03-22 & 22:43 & 22:45 & 23:17 & X9.4 & S26E28 & 6555 & Yes \\
32 & 1991-03-25 & 07:58 & 08:18 & 08:44 & X5.3 & S24W13 & 6555 & Yes \\
33 & 1991-06-01 & 15:09 & 15:29 & 16:14 & X12.0 & N25E90 & 6659 & Yes \\
34 & 1991-06-04 & 03:37 & --- & 07:30 & X12.0 & N30E70 & 6659 & Yes \\
35 & 1991-06-06 & 00:54 & 01:12 & 01:35 & X12.0 & N30E44 & 6659 & Yes \\
36 & 1991-06-09 & 01:37 & 01:40 & 03:04 & X10.0 & N33E04 & 6659 & Yes \\
37 & 1991-06-11 & 02:09 & 02:29 & 03:20 & X12.0 & N31W17 & 6659 & Yes \\
38 & 1991-06-15 & 06:33 & 07:51 & 09:17 & X12.0 & N33W69 & 6659 & Yes \\
39 & 1991-10-27 & 05:38 & 05:49 & 06:18 & X6.1 & S13E15 & 6891 & Yes \\
40 & 1992-11-02 & 02:31 & 03:08 & 03:28 & X9.0 & S26W87 & 7321 & Yes \\
41 & 1997-11-06 & 11:49 & 11:55 & 12:01 & X9.4 & S18W63 & 8100 & Yes \\
42 & 2000-07-14 & 10:03 & 10:24 & 10:43 & X5.7 & N22W07 & 9077 & Yes \\
43 & 2001-04-02 & 21:32 & 21:51 & 22:03 & X20.0 & N19W73 & 9393 & Yes \\
44 & 2001-04-06 & 19:10 & 19:21 & 19:31 & X5.6 & S21E31 & 9415 & Yes \\
45 & 2001-04-15 & 13:19 & 13:50 & 13:55 & X14.4 & S20W85 & 9415 & Yes \\
46 & 2001-08-25 & 16:23 & 16:45 & 17:04 & X5.3 & S17E34 & 9591 & No \\
47 & 2001-12-13 & 14:20 & 14:30 & 14:35 & X6.2 & N16E09 & 9733 & No \\
48 & 2003-10-23 & 08:19 & 08:35 & 08:49 & X5.4 & S21E88 & 10486 & Yes \\
49 & 2003-10-28 & 09:51 & 10:30 & 10:44 & X17.2 & S16E08 & 10486 & Yes \\
50 & 2003-10-29 & 20:37 & 20:49 & 21:01 & X10.0 & S15W02 & 10486 & Yes \\
51 & 2003-11-02 & 17:03 & 17:25 & 17:39 & X8.3 & S14W56 & 10486 & Yes \\
52 & 2003-11-04 & 19:29 & 19:50 & 20:06 & X28.0 & S19W83 & 10486 & Yes \\
53 & 2005-01-20 & 06:36 & 07:01 & 07:26 & X7.1 & N14W61 & 10720 & Yes \\
54 & 2005-09-07 & 17:17 & 17:40 & 18:03 & X17.0 & S11E77 & 10808 & Yes \\
55 & 2005-09-08 & 20:52 & 21:06 & 21:17 & X5.4 & S12E75 & 10808 & Yes \\
56 & 2005-09-09 & 19:13 & 20:04 & 20:36 & X6.2 & S12E67 & 10808 & Yes \\
57 & 2006-12-05 & 10:18 & 10:35 & 10:45 & X9.0 & S07E68 & 10930 & Yes \\
58 & 2006-12-06 & 18:29 & 18:47 & 17:00 & X6.5 & S05E64 & 10930 & Yes \\
59 & 2011-08-09 & 07:48 & 08:05 & 08:08 & X6.9 & N17W69 & 11263 & No \\
60 & 2012-03-07 & 00:02 & 00:24 & 00:40 & X5.9 & N11E27 & 11429 & Yes \\
61 & 2017-09-06 & 11:53 & 12:02 & 12:10 & X9.3 & S08W33 & 12673 & Yes \\
62 & 2017-09-10 & 15:35 & 16:06 & 16:31 & X8.2 & S08W88 & 12673 & Yes \\

\end{longtable}
\end{center}

\subsection{GLE events caused by SARs}
\label{subsect:dataGLE}

According to the GLE events and the SARs, the GLE events and the corresponding ARs are listed in Table \ref{tab:02}. In the Table, the No. of the GLE event in column 1, date in column 2, the source location of the GLE event in column 3, the flare associated with the GLE event in column 4, the NOAA number of the AR in column 5, whether the AR is a SAR in column 6. We can see from Table \ref{tab:02} that there were 46 GLEs during solar cycles 21-24. Of the 46 GLEs, 25 GLEs were caused by SARs, namely that 54.3\% of the GLE events during solar cycles 21-24 were caused by SARs.

\begin{center}
\begin{longtable}{cccccc}
\caption[]{The GLE events caused by SARs during solar cycles 21-24.}\label{tab:02} \\
  \hline
  GLE No. & Date & Location & Flare & AR & SAR? \\
      &~yyyy-mm-dd~ & & & & \\
  \hline
\endfirsthead
\multicolumn{6}{l}%
{{continued from previous page}} \\
  \hline
  GLE No. & Date & Location & Flare & AR & SAR? \\
      &~yyyy-mm-dd~ & & & & \\
  \hline
\endhead

\hline \multicolumn{6}{r}{{continued on next page}} \\
\endfoot
\hline
\endlastfoot

27 & 1976-04-30 & S08W46 & 2B/X2.0 & 700 & No \\
28 & 1977-09-19 & N08W57 & 3B/X2.0 & 889 & No \\
29 & 1977-09-24 & N10W120 & --- & 889 & No \\
30 & 1977-11-22 & N24W40 & 2B/X1.0 & 939 & No \\
31 & 1978-05-07 & N23W72 & 2B/X2.0 & 1095 & No \\
32 & 1978-09-23 & N35W50 & 3B/X1.0 & 1294 & No \\
34 & 1981-04-10 & N07W36 & 2B/X2.3 & 3025 & No \\
35 & 1981-05-10 & N03W75 & 2B/M1.3 & 3079 & No \\
36 & 1981-10-12 & S18E31 & 2B/X3.1 & 3390 & Yes \\
37 & 1982-11-26 & S12W87 & 2B/X4.5 & 3994 & No \\
38 & 1982-12-07 & S19W86 & 1B/X2.8 & 4007 & No \\
39 & 1984-02-16 & S-W130 & --- & 4408 & No \\
40 & 1989-07-25 & N26W85 & 1B/X2.5 & 5603 & No \\
41 & 1989-08-16 & S15W85 & 2N/X20 & 5629 & Yes \\
42 & 1989-09-29 & S24W105 & 1B/X9.8 & 5698 & Yes \\
43 & 1989-10-19 & S25E09 & 3B/X13 & 5747 & Yes \\
44 & 1989-10-22 & S27W32 & 1N/X2.9 & 5747 & Yes \\
45 & 1989-10-24 & S29W57 & 2N/X5.7 & 5747 & Yes \\
46 & 1989-11-15 & N11W28 & 2B/X3.2 & 5786 & No \\
47 & 1990-05-21 & N34W37 & 2B/X5.5 & 6063 & Yes \\
48 & 1990-05-24 & N36W78 & 1B/X9.3 & 6063 & Yes \\
49 & 1990-05-26 & N35W103 & --/X1.4 & 6063 & Yes \\
50 & 1990-05-28 & N35W120 & C9.7 & 6063 & Yes \\
51 & 1991-06-11 & N32W15 & 2B/X12 & 6659 & Yes \\
52 & 1991-06-15 & N36W70 & 2B/X12 & 6659 & Yes \\
53 & 1992-06-25 & N09W69 & 2B/X3.9 & 7205 & No \\
54 & 1992-11-02 & S25W100 & --/X9.0 & 7321 & Yes \\
55 & 1997-11-06 & S18W68 & 2B/X9.4 & 8100 & Yes \\
56 & 1998-05-02 & S15W15 & 3B/X1.1 & 8210 & No \\
57 & 1998-05-06 & S11W65 & 1N/X2.7 & 8210 & No \\
58 & 1998-08-24 & N18E09 & 3B/M7.1 & 8307 & Yes \\
59 & 2000-07-14 & N22W07 & 3B/X5.7 & 9077 & Yes \\
60 & 2001-04-15 & S20W85 & 2B/X14 & 9415 & Yes \\
61 & 2001-04-18 & S20W115 & --/C2.2 & 9415 & Yes \\
62 & 2001-11-04 & N06W18 & 3B/X1.0 & 9684 & No \\
63 & 2001-12-26 & N08W54 & 1B/M7.1 & 9742 & No \\
64 & 2002-08-24 & S02W81 & 1F/X3.1 & 10069 & Yes \\
65 & 2003-10-28 & S16E08 & 4B/X17.0 & 10486 & Yes \\
66 & 2003-10-29 & S15W02 & 2B/X10.0 & 10486 & Yes \\
67 & 2003-11-02 & S14W56 & 2B/X8.3 & 10486 & Yes \\
68 & 2005-01-17 & N15W25 & 3B/X3.8 & 10720 & Yes \\
69 & 2005-01-20 & N14W61 & 2B/X7.1 & 10720 & Yes \\
70 & 2006-12-13 & S06W23 & 4B/X3.4 & 10730 & Yes \\
71 & 2012-05-17 & N11W76 & 1F/M5.1 & 11476 & No \\
72 & 2017-09-10 & S08W88 & --/X8.2 & 12673 & Yes \\

\end{longtable}
\end{center}

\subsection{Super geomagnetic storms caused by SARs}
\label{subsect:dataGeoStorm}

The ARs that produced super geomagnetic storms (SGSs) during different solar cycles have been investigated by many researchers \citep[e.g.][]{Cliver+Crooker+1993, Zhang+etal+2007}. \cite{Meng+etal+2019} collected various information on the ARs that produced CMEs responsible for the super geomagnetic storms during solar cycles 19-24. According to the ARs related to the super geomagnetic storms during solar cycles 21-24, each super geomagnetic storm and the corresponding AR during solar cycles 21-24 is listed in Table \ref{tab:03}. In the Table \ref{tab:03}, column 1 is the number of the super geomagnetic storm, column 2 is the date, column 3 is the super geomagnetic storm intensity, the source location of the AR in column 4, the NOAA number of the AR in column 5, whether the AR is a SAR in column 6. We can see from Table \ref{tab:03} that there are 24 super geomagnetic storms during solar cycles 21-24. Of the 24 super geomagnetic storms, 12 of them were caused by SARs, namely 50\% of the super geomagnetic storms were caused by SARs.

\begin{center}
\begin{longtable}{cccccc}
\caption[]{The SGSs and the related ARs during solar cycles 21-24.}\label{tab:03} \\
  \hline
  No. & Date & Dst & Location & AR & SAR? \\
      &~yyyy-mm-dd~ &~nT~& & & \\
  \hline
\endfirsthead
\multicolumn{6}{l}%
{{continued from previous page}} \\
  \hline
  No. & Date & Dst & Location & AR & SAR? \\
      &~yyyy-mm-dd~ & & & & \\
  \hline
\endhead

\hline \multicolumn{6}{r}{{continued on next page}} \\
\endfoot
\hline
\endlastfoot

1 & 1981-04-13 & -311 & N07W36 & 3025 & No \\
2 & 1982-07-14 & -325 & N11E36 & 3804 & Yes \\
3 & 1982-09-06 & -289 & N12E35 & 3886 & No \\
4 & 1986-02-09 & -307 & S11W21 & 4711 & No \\
5 & 1986-02-09 & -307 & N32E22 & 5395 & Yes \\
6 & 1989-03-14 & -589 & N23W24 & 5687 & No \\
7 & 1989-09-19 & -255 & S25E09 & 5747 & Yes \\
8 & 1989-10-21 & -268 & N11W28 & 5786 & No \\
9 & 1989-11-17 & -266 & N24E28 & 6007 & No \\
10 & 1990-04-10 & -281 & S26E28 & 6555 & Yes \\
11 & 1991-03-25 & -298 & S13E15 & 6891 & Yes \\
12 & 1991-10-29 & -254 & S14W20 & 6909 & No \\
13 & 1991-11-09 & -354 & S26E07 & 7154 & No \\
14 & 1992-05-10 & -288 & N15W66 & 8933 & No \\
15 & 2000-04-07 & -288 & N22W07 & 9077 & Yes \\
16 & 2000-07-16 & -301 & N20W19 & 9393 & Yes \\
17 & 2001-03-31 & -387 & S23W09 & 9415 & Yes \\
18 & 2001-04-11 & -271 & N06W18 & 9684 & No \\
19 & 2001-11-06 & -292 & S15W02 & 10486 & Yes \\
20 & 2003-10-30 & -383 & S16E08 & 10486 & Yes \\
21 & 2003-10-30 & -353 & S01E16 & 10501 & No \\
22 & 2003-11-20 & -422 & N09E05 & 10501 & No \\
23 & 2004-11-08 & -374 & N09W17 & 10696 & Yes \\
24 & 2004-11-10 & -263 & N09W17 & 10696 & Yes \\

\end{longtable}
\end{center}

\subsection{Extreme space weather events caused by SARs during different solar cycles}
We use N$_{\textrm{SAR1}}$, N$_{\textrm{SAR2}}$ to indicate the numbers of SARs that can produce and can not produce extreme space weather events during a solar cycle, respectively. N$_{\textrm{SAR}}$ is used to indicate the total number of the SARs during a solar cycle. The extreme space weather events caused by SARs during different solar cycles (SCs) were analyzed and the derived results were shown in Table \ref{tab:04}. We can see from Table \ref{tab:04}, only 29 SARs can produce $\ge$X5.0 flares. The numbers of SARs that can produce $\ge$X5.0 flares in solar cycles 21, 22, 23 and 24 are 8, 11, 8 and 2, respectively, and the numbers of $\ge$X5.0 flares caused by SARs in solar cycles 21-24 are 13, 21, 15 and 3, respectively. The numbers of SARs that can produce GLE events in solar cycles 21, 22, 23 and 24 are 1, 6, 7 and 1, respectively, and the numbers of GLE events caused by SARs in solar cycles 21-24 are 1, 12, 11 and 1, respectively. Only 10 SARs can produce super geomagnetic storms. The numbers of SARs that can produce super geomagnetic storms in solar cycles 21, 22, 23 and 24 are 1, 4, 5 and 0, respectively, and the numbers of super geomagnetic storms caused by SARs in solar cycles 21-24 are 1, 4, 7 and 0, respectively. The results indicate that the contribution to the extreme space weather events made by SARs in Solar Cycle 24 is the smallest. Most of extreme space weather events caused by the SARs, especially for GLE events and super geomagnetic storms, appeared in solar cycles 22 and 23.

\begin{table}
\bc
\begin{minipage}[]{110mm}
\caption[]{Extreme space weather events caused by SARs during different solar cycles\label{tab:04}}\end{minipage}
\setlength{\tabcolsep}{1pt}
\small
 \begin{tabular}{ccccccccccccc}
  \hline
   SC & \multicolumn{4}{c}{SARs and $\ge$X5.0 flares} & \multicolumn{4}{c}{SARs and GLE events} & \multicolumn{4}{c}{SARs and Dst$\le$-250 nT storms} \\
  \cline{2-13}
   & N$_{\textrm{SAR1}}$ & N$_{\textrm{SAR2}}$ & N$_{\textrm{SAR}}$ & N$_{\ge \textrm{X5}}$ & N$_{\textrm{SAR1}}$&N$_{\textrm{SAR2}}$ & N$_{\textrm{SAR}}$ & N$_{\textrm{GLE}}$ & N$_{\textrm{SAR1}}$ & N$_{\textrm{SAR2}}$ & N$_{\textrm{SAR}}$ & N$_{\textrm{SGS}}$ \\
  \hline
  21 & 8 & 9 & 17 & 13 & 1 & 16 & 17 & 1 & 1 & 16 & 17 & 1\\
  22 & 11 & 5 & 16 & 21 & 6 & 10 & 16 & 12 & 4 & 12 & 16 & 4\\
  23 & 8 & 4 & 12 & 15 & 7 & 5 & 12 & 11 & 5 & 7 & 12 & 7\\
  24 & 2 & 4 & 6 & 3 & 1 & 5 & 6 & 1 & 0 & 6 & 6 & 0\\
total & 29 & 22 & 51 & 51 & 15 & 36 & 51 & 25 & 10 & 41 & 51 & 12\\
  \hline
\end{tabular}
\ec
\end{table}

\subsection{The properties of the source locations of the extreme space weather events caused by SARs}

\begin{figure}
   \centering
  \includegraphics[width=0.85\textwidth, angle=0]{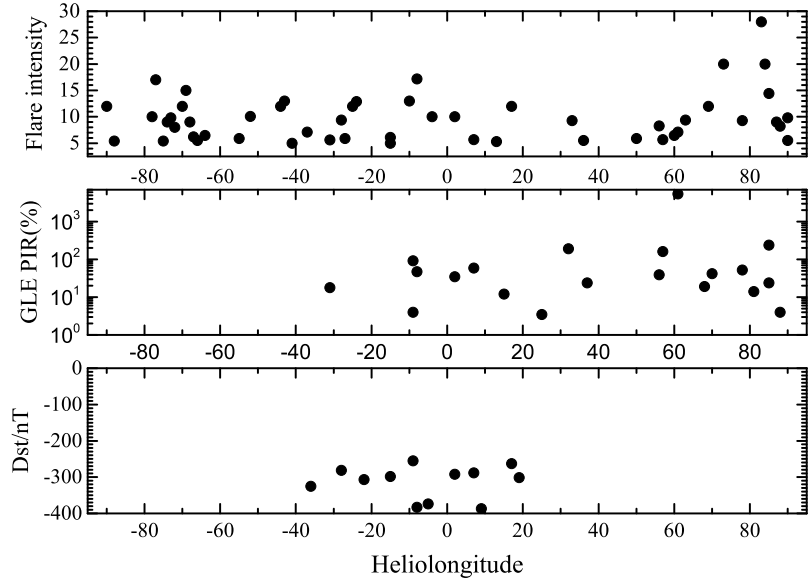}
   \caption{The heliolongitudinal distribution of the source locations of extreme space weather events caused by SARs during solar cycles 21-24. From the top to bottom, it shows the $\ge$X5.0 flares, peak increase rate (PIR) of the GLEs and super geomagnetic storms, respectively.}
   \label{fig:01}
\end{figure}

The longitudinal distribution of the source locations of the extreme space weather events caused by SARs are shown in Figure \ref{fig:01}. As shown in the top panel of Figure \ref{fig:01}, the longitudinal scope for the flares with intensities $\ge$X5.0 caused by SARs ranged from E90 to W90. The longitudinal area of the source locations of GLE events caused by SARs ranged from E31 to W120 according to Table \ref{tab:02}, and the heliolongitude of the strongest GLE events is located at around W60, which is shown in the second panel of Figure \ref{fig:01}. To be noted that the abscissa scope in the second panel only ranged from E90 to W90 to be consistent with the upper and lower panels. The peak increase rate (PIR) for each GLE event that occurred during solar cycles 21-23 is obtained from the article by \cite{Belov+etal+2010}, while the PIR for the GLE that occurred on 10 September 2017 is obtained from the article by \cite{Zhao+etal+2018}. The longitudinal span of the source locations of the super geomagnetic storms caused by SARs ranged from E36 to W19, which is shown in the lowest panel of Figure \ref{fig:01}, indicating that only the CMEs produced by the corresponding SARs with source locations around solar disk center can produce super geomagnetic storms.

\section{Discussion}
\label{sect:discussion}

\begin{figure}
   \centering
  \includegraphics[width=9.5cm, angle=90]{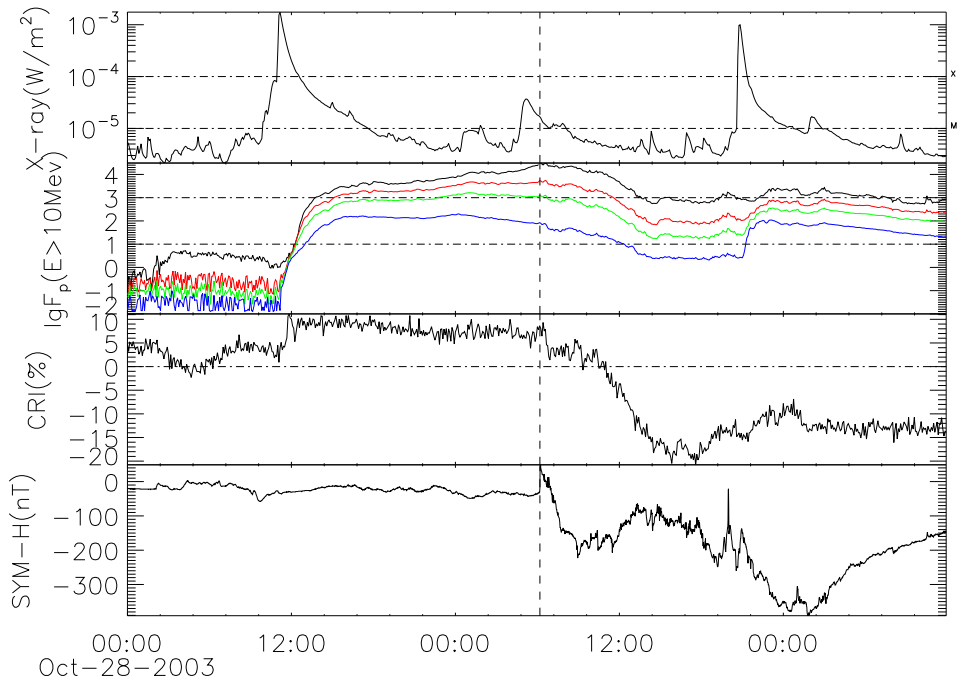}
   \caption{The solar activities and their geoeffectiveness caused by SAR 10486 during 28-30, October 2003. From the top to bottom, it shows solar flares, the fluxes of E$>$10, 30, 50 and 100 MeV protons observed by GOES 10, the counting rate of cosmic rays observed by Oulu neutron monitor and the SYM-H index, respectively. The dashed line indicated the moment of the SSC.}
   \label{fig:02}
\end{figure}

Among the 51 SARs, 18 of them did not produce a $\ge$X5.0 flare, nor did they produce a GLE event or a super magnetic storm. For the remaining 33 SARs, each of them produced at least an extreme space weather event. In this context, 64.7\% (or 33/51) of the SARs can produce extreme space weather events. 10 SARs produced both a $\ge$X5.0 flare and a GLEs, but they did not produced a super geomagnetic storm. Only 4 SARs not only produced at least a $\ge$X5.0 flare, but also produced at least a GLE event and a super geomagnetic storm (SGS). Here, we give an example shown in Figure \ref{fig:02}. As shown in Figure \ref{fig:02}, the SAR 10486 with source location at S16E08 produced a X17.2 flare and a CME with projected speed 2459 km/s on 28 October 2003, the flux of E$>$10 MeV protons increased quickly after the flare and the CME, and reached its peak flux 29500 pfu at 06:10 UT, 29 October 2003, which is consistent with the moment of sudden storm commencement (SSC), indicating that the flux of E$>$10 MeV protons reached its peak flux at the moment when the CME-driven shock reached the magnetosphere. As shown in the second panel of Figure 2 that the cosmic ray intensity (CRI) increased obviously, namely that a GLE event was observed. When the CME-driven shock and the CME itself reached the magnetosphere, it triggered a super geomagnetic storm (Dst$_{min}$=-353 nT, SYM-H$_{min}$=-391 nT). The super geomagnetic storm was mainly caused by the ICME \citep{Zhang+etal+2008}. SAR 10486 produced X5.4+X17.2+X10 +X8.3+X28 flares, 3 GLE events and 2 super geomagnetic storms. 4 SARs, that can produce not only a $\ge$X5.0 flare, but also a GLE event and a super geomagnetic storm, were listed in Table \ref{tab:05}.

\begin{table}
\bc
\begin{minipage}[]{110mm}
\caption[]{SARS producing all three types of extreme space weather events\label{tab:05}}\end{minipage}
\setlength{\tabcolsep}{1pt}
\small
 \begin{tabular}{cccc}
  \hline
  SAR & $\ge$X5.0 flare & No. of GLE & No. of SGS \\
  \hline
  5747 & X13+X5.7 & 3 & 1\\
  9077 & X5.7 & 1 & 1\\
  9415 & X5.6+X14.4 & 2 & 1\\
  10486 & X5.4+X17.2+X10 +X8.3+X28 & 3 & 2\\
  \hline
\end{tabular}
\ec
\end{table}

\begin{table}
\bc
\begin{minipage}[]{110mm}
\caption[]{The comparison of different SARs during Solar Cycle 22\label{tab:06}}\end{minipage}
\setlength{\tabcolsep}{1pt}
\small
\begin{minipage}{\textwidth}
\let\footnoterule\relax
 \begin{tabular}{ccccccccc}
  \hline
SAR & Date on the disk & LA & FI & X-class flares & 10.7cm peak flux & SPEI\footnote{Solar Proton Event Intensity.} & Ap & Authors\\
 &  & ($\mu h$) &  &  & (s.f.u.) & (pfu) &  & \\
\hline
5800 & 891119-1202 & 590 & 3.6 & X1.0+X2.6 & 2100 & 7300 & 110 & Tian et al. (2002)\\
6022 & 900414-0427 & 1070 & 1.4 & X1.4 & 11000 & 12 & 125 & Tian et al. (2002)\\
6703 & 910628-0713 & 280 & 1.9 & X1.9 & 1778 & 2300 & 135 & Tian et al. (2002)\\
7154 & 920504-0515 & 500 & 2.1 & (M7.4)$<$X1.0 & 3100 & 4600 & 180 & Tian et al. (2002)\\
7671 & 940213-0226 & 450 & 1.7 & (M4.0)$<$X1.0 & 190 & 10000 & 95 & Tian et al. (2002)\\
5312 & 890106-0120 & 1800 & 20.64 & 2(X1.1+X1.4)+X2.3+X2.1 & 1400 & NG\footnote{Not Given.}  & NG  & Chen et al. (2011)\\
5533 & 890609 & 920 & 11.37 & X4.1+X3.0 & 1100 &NG  & NG & Chen et al. (2011)\\
5669 & 890829-0912 & 3080 & 13.32 & X1.2+X1.1+X1.3 & 4800 &NG  &NG & Chen et al. (2011)\\
5852 & 891225-1231 & 1500 & 6.42 & X2.8 & 1600 & NG &NG & Chen et al. (2011)\\
6471 & 910125-0208 & 2210 & 15.27 & X10+X1.9 & 3500 &NG  &NG  & Chen et al. (2011)\\
6538 & 910305-0317 & 910 & 17.08 & X1.5+X2+X5.5+X2.5+X1.7 & 3500 &NG   &NG   & Chen et al. (2011)\\
6545 & 910311-0322 & 830 & 16.93 & X1.7+X1.3+X3.9+X1.8+X1.8+X1.0 & 3600 &NG  &NG  & Chen et al. (2011)\\
  \hline
\end{tabular}
\end{minipage}
\ec
\end{table}

Different lists of SARs given by different researchers will lead to different extreme space weather events caused by the corresponding SARs. Which one is the better? To answer the question, we made a comparison between the SARs given by \cite{Tian+etal+2002} with those given by \cite{Chen+etal+2011} in Solar Cycle 22. There were 14 SARs during Solar Cycle 22 given by \cite{Tian+etal+2002}, while 16 SARs in Solar Cycle 22 given by \cite{Chen+etal+2011}. 9 SARs occurred in both lists of the SARs, so we only compare the different SARs in the two lists. The comparison is shown in Table \ref{tab:06}. We can see from Table \ref{tab:06} that the solar flare activities of the SARs proposed by \cite{Chen+etal+2011} shown in Table \ref{tab:06} were much stronger than those proposed by \cite{Tian+etal+2002}. The comparison between the different SARs during Solar Cycle 22 by \cite{Tian+etal+2002} and by \cite{Chen+etal+2011} implies that the concept of SAR proposed by \cite{Chen+etal+2011} puts more emphasis on flare activity than that proposed by \cite{Tian+etal+2002}, while the concept of SAR proposed by \cite{Tian+etal+2002} paid more attention to the geoeffectiveness of SARs than that proposed by \cite{Chen+etal+2011}. The comparison tell us that both criteria proposed by \cite{Tian+etal+2002} and by \cite{Chen+etal+2011} should be improved. Anyway, there was only a complete list of the SARs for solar cycles 21-24 according to the criteria proposed by \cite{Chen+etal+2011}. This is the reason why we study the extreme space weather events during solar cycles 21-24 caused by SARs only based on the 51 SARs according to the criteria proposed by \cite{Chen+etal+2011}. The criteria for SAR will be more reasonable after more study and the extreme space weather events caused by SARs will be revised.

\section{Summary}
\label{sect:summary}

The following gives the major points concluded from the study:

(i)  There were 62 $\ge$X5.0 flares and 51 SARs during solar cycles 21-24. Of the 62 $\ge$X5.0 flares, 51 of them were produced by SARs, namely that 83.9\% of the $\ge$X5.0 flares were produced by SARs. Of the 51 $\ge$X5.0 flares, the numbers of X5.0-X9.9 and $\ge$X10.0 flares are 41 and 21, respectively, and 78.05\% of the X5.0-X9.9 and 95.24\% of the $\ge$X10.0 solar flares were produced by SARs, respectively. The number of $\ge$X5.0 flares produced by the SARs in solar cycles 21, 22, 23 and 24 were 13, 21, 15 and 3, respectively. Only 29 SARs can produce $\ge$X5.0 flares, indicating that only 56.9\% of the SARs can produce $\ge$X5.0 flares. The longitudinal area of the source locations of the flares with intensities $\ge$X5.0 caused by SARs ranged from E90 to W90.

(ii) 46 GLEs registered during solar cycles 21-24. Of the 46 GLE events, 25 GLE events were caused by SARs, namely that 54.3\% of the GLEs were caused by SARs. The numbers of GLE events caused by the SARs in solar cycles 21, 22, 23 and 24 were 1, 12, 11 and 1, respectively, indicating  that most of the GLE events caused by the SARs came form solar cycles 22 and 23. Only 15 SARs can produce GLE events, namely only 29.4\% of the SARs can produce GLE events. The longitudinal scope of the source locations of GLE events caused by SARs ranged from E31 to W120. The longitude of the source location for the strongest GLE event is located around W60.

(iii) There were 24 super geomagnetic storms during solar cycles 21-24. 12 super geomagnetic storms were caused by SARs, namely 50\% of the super geomagnetic storms were caused by SARs. The numbers of super geomagnetic storms caused by the SARs in solar cycles 21, 22, 23 and 24 were 1, 4, 7 and 0, respectively. Only 10 SARs can produce super geomagnetic storms, indicating that only 19.6\% of the SARs can produce super geomagnetic storms. The longitudinal span of the source locations of super geomagnetic storms caused by SARs ranged from E36 to W19.

(iv) Of the 51 SARs, only 33 SARs can produce at least one extreme space weather event, while none of the rest 18 SARs can produce an extreme space weather event. There were only 4 SARs, each of them can produce not only  a $\ge$X5.0 flare, but also a GLE event and a super geomagnetic storm. Most of the extreme space weather events caused by the SARs appeared during solar cycles 22 and 23, especially for the GLE events and super geomagnetic storms. Solar Cycle 24 is a very weak cycle, the number of the SARs is small and the number of extreme space weather events caused by the SARs is also small, especially that there was no super geomagnetic storm in Solar Cycle 24.

\normalem
\begin{acknowledgements}

We are very grateful to the anonymous referee for her/his review of our manuscript and for the helpful suggestions. We thank NOAA for providing the solar soft x-ray data and the fluxes of protons with different energies, and the Center for Geomagnetism and Space Magnetism, Kyoto University, for providing the Dst and SYM-H index, and Oulu cosmic ray station providing cosmic ray intensity data, and Institute of Geophysics, China Earthquake Administration for providing SSC time. This work was supported by the National Natural Science Foundation of China (Grant Nos. 41774085, 41074132, 41274193, 41474166, 41774195, and 41874187).

\end{acknowledgements}

\label{lastpage}

\end{document}